# An Interpretation of Quantum Mechanics based on a Waveguide Analogy

## Roald Ekholdt [*]

**Abstract:** Based on an observation that the basic mode of a common microwave waveguide is a solution of the Klein-Gordon equation quantum mechanics is modeled as the wave-function propagated inside a waveguide. The guide width is determined by the rest-energy and the potential energy. Moreover, a geometrical optics model of the propagation illustrates how the width as a function of distance determines the velocity and accelerations of the particle. The model entails the disappearance of the principle of superposition of the wave-function – and with it action-at-a-distance, and that interactions are conveyed via the potential, i.e. by photons.

## 1. Introduction

The conceptual abyss between the Bohr-Sommerfeld atomic model and the Schrödinger wave mechanics theory is an enigma in the history of quantum theory. Especially after Sommerfeld's introduction of relativistic mass accounted for the fine-structure in the spectral lines this classical mechanics particle theory seemed to be successfully on the right track – until de Broglie's introduced his wave-particle hypothesis. Bohr introduced the principle of correspondence between the theories, without giving an explanation of underlying phenomena.

Starting out from de Broglie's quantum wave hypothesis this paper introduces a waveguide model for quantum mechanics. It is based on the observation that the Klein-Gordon equation[1], which Schrödinger type equations approximate in the low frequency and low group velocity range, governs the basic propagation mode of the common rectangular microwave waveguide. This waveguide analogy represents the relativistic transforms of both the frequency of the wave and the clock, which de Broglie based his theory upon.[2]

We note, however, the difference between Schrödinger's concept of standing waves, or vibrations, in his model of the hydrogen atom and de Broglie's concept of traveling waves. Schrödinger makes this difference clear in his first paper on wave mechanics, where he introduced his time-independent equation:

'The main difference is that de Broglie thinks of progressive waves, while we are led to stationary proper vibrations if we interpret our formulae as representing vibrations.' [3]

On Schrödinger's theory de Broglie wrote [4]:

'At that epoch I read Schrödinger's papers with great admiration, while reflecting much on their contents. On three points, however, I did not feel to be in agreement with the eminent Austrian physicist. Firstly, the wave equation that he attributed to the wave ψ was not relativistic and I was too strongly convinced of the existence of a close connection between the Theory of Relativity and Wave Mechanics to be satisfied by a non-relativistic wave equation; but that difficulty was swiftly removed since, from July 1926, several authors, including myself, found a form of wave equation, which now is known as the Klein-Gordon equation, of which Schrödinger's is a degenerate form that corresponds to the Newtonian approximation. Another point where my views were not in accord with Schrödinger's was that he, while retaining the idea that the wave ψ in the physical space is a real wave, seems to abandon completely the idea of the localisation of the particle in the wave, which was not in accordance with my initial conceptions. Finally, while recognising that considering a wave in the configuration space constitutes a very useful formalism for the prediction of the properties of an ensemble of interacting particles, I considered as certain that the movement of the diverse particles and their propagation operated in the physical space and time.'

---

[*] Retired. Formerly with the Norwegian Defence Research Establishment, the Norwegian Telecom Research Institute and the Norwegian Post and Telecommunications Authority. E-mail: roaldo@ieee.org. Address: Granstubben 2, N-2020 Skedsmokorset, Norway.

Had de Broglie's wave-particle project not collapsed at the 1927 Solvay conference [5] quantum mechanics might eventually have taken a course in the direction proposed in the present paper. That one should find the waveguide solution directly by mathematical methods, however, seems unlikely, so this solution could probably first have appeared in 1936 when the microwave waveguide was invented.[6] R.P. Feynman briefly mentions the analogy in a lecture on microwave waveguides, but he seems not to have elaborated on it later.[7]

The main result of this paper is that the quantum mechanics of a single electron at non-relativistic velocities may be treated in the following steps:
- Use classical mechanics as a first approximation to establish a trajectory.
- Convert the classical trajectory to a waveguide structure which confines the wave; while the interactions between electrons are conveyed by photons only – not by the wave.
- Use the waveguide model to introduce such wave aspects as the tunnel effect and the stationary orbits in the Bohr- Sommerfeld planetary trajectory model.
- Introduce wave based corrections to the classical potential – i.e. a local quantum potential.

## 2. A derivation of the Schrödinger equation directly from the de Broglie relativity relations

In 1924 de Broglie presented the theory of the electron as a dual wave - particle entity based on an analogy with the Planck-Einstein electromagnetic quantum — the photon, and the special theory of relativity, arriving at the rest energy and momentum relations $E_o = m_o c^2 = \hbar \omega_o$ and $P = \hbar k$.[8]

Contrafactually, de Broglie could have derived a wave equation from Einstein's relativistic energy equation

$$E_{rel}^2 = (m_o c^2)^2 + (cP)^2 \qquad (1)$$

by inserting his relations, which results in the dispersion equation

$$(\hbar \omega)^2 = (\hbar \omega_o)^2 + (c \hbar k)^2 \qquad (2)$$

The relation between group velocity and phase velocity is thus

$$v_g = \frac{\partial \omega}{\partial k} = c^2 \frac{k}{\omega} = \frac{c^2}{v_{ph}} \qquad (3)$$

Further, assuming as an elementary solution the wave function $\psi = A e^{i(\omega t - kx)}$, he could have obtained the corresponding wave equation

$$\hbar \frac{\partial^2}{\partial t^2} \psi = (\hbar \omega_o)^2 \psi - (c\hbar)^2 \frac{\partial^2}{\partial x^2} \psi \qquad (4)$$

This equation, the so-called Klein-Gordon equation was, however, first presented in 1926 as a relativistic development of Schrödinger's equation.

Moreover, de Broglie could have added a potential to the $\hbar\omega_o$ term. As a next step it would be natural to do the same also in the case of a varying potential, although the transform from dispersion relation to wave equation would only be valid for a position-independent potential. Clearly, de Broglie would have based his equation on progressing waves in ordinary 3-dimensional space instead of Schrödinger's standing waves in configuration space.

A Schrödinger type equation could have been obtained directly by introducing the low group velocity approximation

$$\hbar\omega = \hbar\omega_o + V(x) + (c\hbar k)^2 / 2\hbar\omega_o \qquad (5)$$

and the corresponding wave equation

$$i\hbar\frac{\partial}{\partial t}\psi = (\hbar\omega_o + V)\psi - (\hbar^2/2m_o)\frac{\partial^2}{\partial x^2}\psi \qquad (6)$$

which we recognize as the Schrödinger equation *as far as form is concerned*. Schrödinger, however, introduced the time dependence for a different purpose, i.e. to account for time-varying potential. Our derivation illustrates that the imaginary unit *i* in the equation results from our choice of functions of the form $\psi = Ae^{i(\omega t - kx)}$; thus, in this context, it does not seem to represent anything fundamentally new, which is often asserted. The operator representation of total energy and linear momentum follows as $H_{operator} = i\hbar\frac{\partial}{\partial t}$ and $P_{operator} = -i\hbar\frac{\partial}{\partial x}$. These operators thus couple the wave concept to the classical mechanics concepts.

Schrödinger probably found the Klein-Gordon equation in his aborted attempt to find a wave equation based on de Broglie's ideas [9], and it seems unfortunate that he did not retain the $\hbar\omega_o$ term in his equations for the non-relativistic velocity range. Surely, he knew about Sommerfeld's use of the relativistic mass variation in his fine-structure theory in his otherwise non-relativistic treatment. [10] Had he retained $\hbar\omega_o$, he might have interpreted $\omega_o$ as a carrier frequency modulated by the frequency $V(x)/\hbar$; but frequency modulation was not known to him at the time — it was patented by E.H.Armstrong in 1933.

### 3. A waveguide analogy

A better physical understanding of both the Klein-Gordon and the Schrödinger equations may be obtained by an analogy with the common rectangular microwave guide; its lowest propagation mode, the transverse electric mode ($TE_{10}$), is represented by the Klein-Gordon equation.

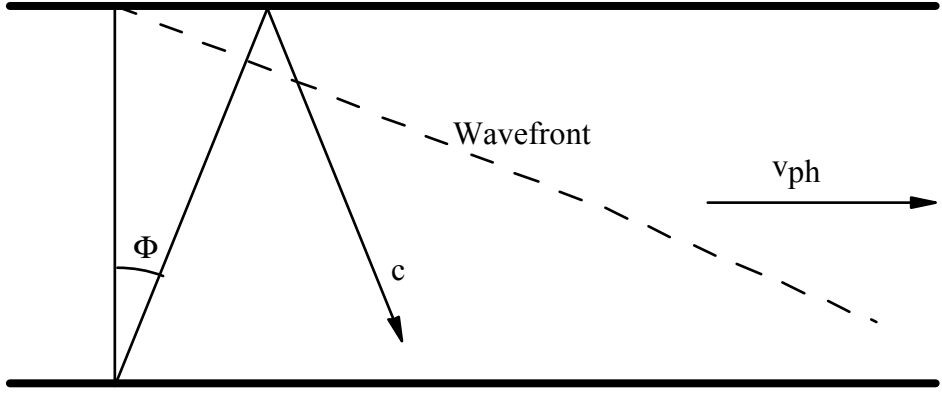

Figure 1.  Top side view of transverse electric, rectangular waveguide: only one of a symmetric pair of plane wave-fronts is shown. (Vertical electric field.)

The $TE_{10}$ mode may be considered as consisting of plane waves that are reflected by the waveguide's metallic walls (where the electric field is zero), which again may be thought of as consisting of rays perpendicular to the phase front of the plane waves.[11] Figure 1. illustrates the trajectory of one such ray. The transmission cut-off occurs when the waveguide width equals a half wavelength -- the rays are then perpendicular to the axial direction. For shorter wavelengths the effective axial energy propagation velocity is $v_g = c \sin\phi$, and the phase velocity is $v_{ph} = c/\sin\phi$. The dispersion equation is thus

$$\omega^2(k) = \omega_o^2(k) + (ck)^2 \qquad (7)$$

where k is the wave number, c the speed of light and $\omega_o$ the cut-off frequency of the mode. We note here that the group velocity also represents the effective axial velocity of a monochromatic wave, as in energy transmission.

From this analogy we may introduce a more basic particle and wave picture: The particle moves along the x-axis and is locked to a ray that propagates at the speed of light in a zigzag pattern. While equation (7) corresponds to the relativistic transform of the wave frequency, the zigzag frequency transforms as the relativistic clock. To begin with we limit our discussion to the rectilinear case; we shall consider the curved case in section 5. The effective velocity of the wave and the particle is a function of the width of the guide, which in the potential-free case is inversely proportional to the rest energy of the electron; the guide width is thus one-half Compton wavelength, i.e.
*h/2mc = 1.21x 10 $^{-12}$* meter.

We may, however, open for varying guide widths as long as the variations are so small per wavelength that no reflections occur — i.e. obeying the Wentzel-Kramer-Brillouin assumption.[12] The addition of an ordinary potential term *V(x)* thus changes the waveguide's cut-off frequency to

$$\omega_{oV} = \omega_o + \frac{V(x)}{\hbar} \qquad (8)$$

The trajectory of an electron may thus be pictured as a waveguide with varying width, corresponding to the addition of a potential, which may be seen as a pressure that is acting on the guide. The mode picture illustrates that the effective velocity depends on the local shape of the waveguide walls, with acceleration and deceleration depending on the gradient, as well as higher derivatives, of the width.

The probability density of finding the particle at a particular position is thus inversely proportional to the group velocity -- as is the case for the intensity of the effective axial wave. The frequency modulation due to the varying potential is thus converted to amplitude modulation. Since we consider a continuous wave we have a picture that opens for a statistical interpretation of the wave function. We may thus introduce a Gibb's ensemble of non-interacting particles belonging to the same wave function, where the actual object under consideration is a *single* particle. It is important to remember that we are only considering a unidirectional wave, while Born's statistical interpretation allows for reflected waves as well.[13] We note that action-at-a-distance, from our perspective, would have to take place between statistically *possible* particles, i.e. fictitious particles, and the single *actual* particle. As pointed out by Nathan Rosen, this aspect is *more bizarre than the non-locality in itself*.[14]

*Instead of a continuous wave we should consider a wave packet in the form of an envelope soliton, which is governed by a non-linear Schrödinger equation.[15] It is important to note that in the single particle case we must interpret the waveguide as existing only in the close vicinity of the particle, while the remainder of the waveguide must be considered as fictitious. The so-called collapse of the wave-function must thus be regarded purely as a mathematical artifice. Moreover, in our waveguide model the wave-function $\Psi$ is confined to the inside of the waveguide, which invalidates the superposition principle for the wave-function – except perhaps in conjunction with Pauli's exclusion principle. The interaction with other particles is conveyed by the potential, i.e. by photons. We note here also that the potential must be considered as representing a statistical photonic field.*

This analysis supports Schrödinger's interpretation that his equation basically represents standing vibrations.

## 4. Square potentials in the waveguide model

The simplest case of variable potential is a concatenation of sections of different but constant potentials. This case also explains the tunnel effect as the coupling by evanescent waves through waveguide sections that are below cut-off for the relevant frequencies, i.e. for imaginary $k$. The effect may be viewed as depending on the way the ray hits the opening to the cut-off section; if it does not hit the particle is reflected. In the case of reflection the returning wave-function must be considered as a new function. This must also be the case when the particle tunnels through. The change of kinetic energy will alter the potential and may thus result in external transmission of photons. These processes may give a clue to the measurement problem – i.e. when the transmission of photons is sufficiently strong to be detectable at the macro level.

## 5. The central potential case

We shall now see that the quantum conditions Bohr determined empirically may be developed from our waveguide model by applying a proposal made by de Broglie. To recapitulate Bohr's theory[16]: Classically, in the circular orbit case the centrifugal force is balanced by the attractive force

$$\frac{mv^2}{r} = \frac{e^2}{r^2} \quad (9)$$

thus

$$mv^2 r = vM = e^2 \quad (10)$$

$M$ represents the angular momentum. It follows that the energy in the hydrogen atom is

$$E = \frac{mv^2}{2} - \frac{e^2}{r} = -\frac{mv^2}{2} \quad (11)$$

To match the Balmer-series he concluded that

$$M = l\hbar \quad (12)$$

where $l$ is the angular quantum number. For the elliptic case Sommerfeld developed an additional radial quantum number.

De Broglie considered[17], in the circular orbit case, that the electron and the phase wave start at the same instant from the same point, but while the electron moves with the velocity $v_g$ the phase wave moves with the much greater phase velocity $v_{ph} = c^2/v_g$. See figure 2.

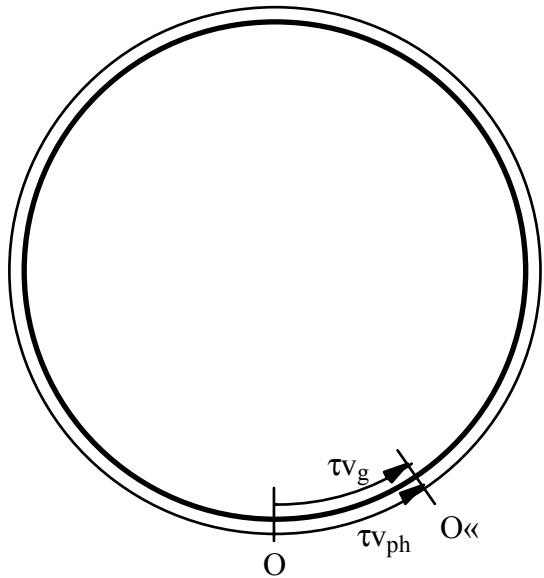

Figure 2. The phase wave overtakes the electron at O' after having moved an additional entire orbit.

The phase wave overtakes the electron at the distance $\tau v_g$ after having moved an additional entire orbit, i.e. a distance of

$$v_{ph}\tau = c^2\tau/v_g = (\tau + T)v_g \tag{13}$$

$T$ is her the time it takes for the particle to make a full orbit.

$$v_g^2(\tau + T)/\tau = c^2 \tag{14}$$

and since $\tau \ll T$

$$v_g^2 T \cong c^2\tau \tag{15}$$

The length of a zigzag period is

$$L = \lambda_o \tan\phi = \lambda_o v_g/c\sqrt{1-(v_g/c)^2} = hv_g/hf_o\sqrt{1-(v_g/c)^2}$$
$$= hv_g/m_o c^2\sqrt{1-(v_g/c)^2} \tag{16}$$

Assuming

$$v_g\tau = nL = nhv_g/m_o c^2\sqrt{1-(v_g/c)^2} \tag{17}$$

where $n$ is an integer; thus

$$\tau = nh/m_o c^2\sqrt{1-(v_g/c)^2} \tag{18}$$

Insertion in (15) yields

$$v_g^2 T = c^2\tau = nh/m_o\sqrt{1-(v_g/c)^2} \tag{19}$$

or

$$m_o v_g^2 T\sqrt{1-(v_g/c)^2} = nh \tag{20}$$

In the non-relativistic case we have $m_o v_g^2 T = M = nh$ – i.e. the same as Bohr's result. De Broglie also obtained the same result, but by invoking a relativistic clock analogy for the particle. Note, however, that the relativistic clock transform is represented in the waveguide analogy. The rest time is proportional to the waveguide width and the moving time is proportional to the time needed for the ray to cross the guide; thus

$$\omega_{rel} = \omega_o \cos\phi = \omega_o \sqrt{1-(v_g/c)^2} \qquad (21)$$

The change between nearest neighboring orbits requires an energy exchange of one photon at the difference frequency. To stimulate such an exchange we may modulate the potential by that frequency, which yet again modulates the width of the waveguide. The moving particle will thus experience width gradients that may cause it to accelerate, or decelerate, depending on the phase. The same process may also occur spontaneously due to the statistical nature of the potential. This process may be considered as related to parametric frequency conversion in electronics and optics.

## 6. Continuously varying potentials

In the case of continuously varying potentials the geometric method breaks down and we must resort to the wave equation and the group velocity concept. Introducing the low velocity approximation

$$v_g \propto (\omega - \omega_o - \omega_{oV})^{1/2} \qquad (22)$$

the probability density is represented by

$$p(x) \propto 1/v_g \propto (\omega - \omega_o - \omega_{oV})^{-1/2} \qquad (23)$$

Since action plays a central role both in quantum theory and the Hamilton-Jacoby equation this equation may show the way to an equation of motion in the quantum case, similar to the de Broglie-Bohm theory, where the concept of the quantum potential was introduced. We follow Bohm's mathematical development.[18] But while he used Schrödinger's equation and its standard interpretation we use our waveguide version.

Essentially, Bohm matched Schrödinger's equation to the Hamilton - Jacobi theory of classical mechanics by introducing the solution in the polar form

$$\psi = R e^{iS/\hbar} \qquad (24)$$

where $R$ and $S$ are real variables. By inserting this function in the Schrödinger equation, Bohm obtained two coupled equations for the solution of $R$ and $S$:

$$\frac{\partial R}{\partial t} = -\frac{1}{2m}(R\nabla^2 S + 2\nabla R \bullet \nabla S) \qquad (25)$$

$$\frac{\partial S}{\partial t} = -[\frac{1}{2m}(\nabla S)^2 + V(x) + U(x)] \qquad (26)$$

where

$$U(x) = -\frac{\hbar^2 \nabla^2 R}{2m_o R} \qquad (27)$$

He then introduced $P(x) = R^2(x)$ as the probability density -- in line with Born's interpretation of the wave function.

He identified (26) as representing the classical Hamilton-Jacobi equation if the term $U(x)$ is omitted, and thus introduced $\nabla S / m$ as velocity. He called the $U(x)$ the quantum potential. *Since it emerges here as a function of the wave function, the problem with action-at-a-distance arises; this phenomenon is inherent in Schrödinger's theory – from our perspective due to the standing wave assumption.* In addition to the action-at-a-distance as such comes the attribute of the quantum potential that it only depends on the shape of *R,* not on its value. *Thus, even the strength of the action does not depend on distance.*

From equation (23), taking into account that we consider a one-dimensional problem, it follows that

$$U(x) = -(\hbar^2/8m_o)\{(\omega - \omega_o - \omega_{oV})^{-1}[\frac{\partial^2 V}{\partial x^2} - \frac{5}{4}(\omega - \omega_o - \omega_V)^{-1}(\frac{\partial V}{\partial x})^2]\} \quad (28)$$

By this approach we obtain a Hamilton-Jacobi equation of motion that is equivalent to a modified Newton's law:

$$m\frac{d^2 x}{dt} = -\frac{\partial}{\partial x}[V(x) + U(x)] \quad (29)$$

This modification does not imply that classical theory applies at the micro level, rather that the classical theory may be derived as an approximation of a quantum theory. The classical tenets of causality and local action only are, however, justified. It has been maintained that the non-appearance of the quantum potential in classical mechanics is due to the small value of the factor $\hbar^2$ in $U(x)$, but from our perspective it is due only to the relatively small potential gradients in classical mechanics applications as compared to quantum mechanics applications.

*This interpretation eliminates the mysterious aspects of U(x): the non-locality as well as the problem of the static electron in a potential well and in the s-states of atoms, which Einstein did not accept.* [19] *Bohm's application of the non-local quantum potential in the dual slit case must in our case be changed to our local quantum potential. In other words, the interference effect is due to electromagnetic interaction between the electron and the two slits.*

J. S. Bell's theory of action-at-a-distance [20] is based on electron spin while all measurements have been performed on photon polarization. [21] This may indicate that spin plays a pivotal role in this effect.

## 7. Dirac's wave equation

Dirac modified Schrödinger's non-relativistic equation into a relativistic equation.[22] He started out from the field-free case and obtained first the Klein-Gordon equation -- in his relativistic momentum operator notation:

$$\{p_o^2 - m^2 c^2 - p_1^2 - p_2^2 - p_3^2\}\psi = 0 \quad (30)$$

But he maintained that it was not possible to construct a positive definite probability function satisfying the second order of $p_o$-operator (or $\partial/\partial t$). Dirac solves this problem by introducing a linear function which transforms in a simple manner under a Lorentz transformation:

$$\{p_o - \alpha_1 p_1 - \alpha_2 p_2 - \alpha_3 p_3\}\psi = 0 \qquad (31)$$

Dirac attributes the $\alpha-$ and $\beta-$ operators to an internal degree of freedom, which he interprets as corresponding to the electron's spin, and introduces the spinor concept.

We shall not go further into Dirac's theory and spin, except for mentioning Penrose's zigzag picture of this theory that resembles our waveguide model.[23] Instead of our picture of a photon pair following the ray, he introduces two particles, Zig and Zag with spin +/-$\hbar$/2 respectively moving with the speed of light. For each reflection they change roles – see figure 3. As in our case the effective velocity is that of the electron.

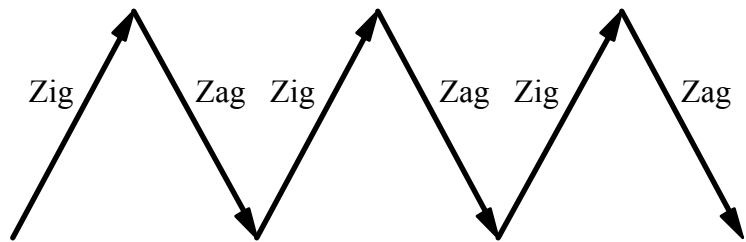

Figure 3. Penrose's zigzag picture of the electron.

## 8. Concluding remarks

The waveguide model demonstrates that the Klein-Gordon equation has a waveguide-like trajectory solution, and that the Schrödinger equation's wave function may be interpreted as confined inside this waveguide. The wave function, however, cannot be viewed as representing a field at the level we consider. The amplitude of the wave function is not defined, although its relative values represent a useful auxiliary statistical function. From our interpretation the stochastic aspects are attributes of the potential, which may be seen as a photonic field. We note, however, that in the many-body case, our assumption of locality requires the use of retarded potentials – but in the stationary case the retardation may cancel out in a statistical treatment.

The wave function of the Klein-Gordon equation may, nevertheless, correspond to a field a deeper level where an explanation of its coupling to the particle and the nature of the electron may be found. One may hope that de Broglie's ideas may conceivably lead to a soliton wave solution.

## Notes